\documentclass[a4paper,12pt]{amsart}
\usepackage{amsfonts,amsmath,amsthm}
\usepackage{amssymb}
\usepackage[square]{natbib}
\newcommand*{\Mbar}{{\mathord{\bar{M}}}}

\newcommand*{\Ahat}{{\hat{A}}}

\newcommand*{\Xhat}{{\hat{X}}}

\newcommand*{\AAA}{{\mathbb{A}}}
\newcommand*{\RR}{{\mathbb{R}}}
\newcommand*{\scrH}{\mathord{\mathcal{H}}}%
\newcommand*{\scrP}{\mathord{\mathcal{P}}}
\newcommand*{\scrS}{\mathord{\mathcal{S}}}

\newcommand*{\Shv}{\mathop{\mathrm{S\lowercase{hv}}}}
\DeclareRobustCommand\openone{\leavevmode\hbox{\small1\normalsize\kern.33em1}}%

\newcommand*{\EsubS}{\ensuremath{\mathord{\mathcal{E}_{\scrS}}}}

\newcommand*{\RsubD}{\ensuremath{\mathord{\RR_{\mathrm{D}}}}}
\newcommand*{\Tr}{\mathop{\mathrm{Tr}}}

\newtheorem{definition}{Definition}
\newtheorem{lemma}[definition]{Lemma}
\newtheorem{proposition}[definition]{Proposition}

\begin{document}
%
\bibliographystyle{apsrev}
\title{\bf Quantum Mechanics as a Space-Time Theory}
\maketitle

\centerline{\large John Corbett\footnote{Mathematics Department,
 Macquarie University, Sydney, NSW 2109, Australia, email: jvc@ics.mq.edu.au}, Thomas Durt\footnote{TENA-TONA
Free University of
Brussels, Pleinlaan 2, B-1050 Brussels, Belgium. email:
thomdurt@vub.ac.be}}¥\bigskip\bigskip

\date{\today}



\date{\today}

\begin{abstract}
We show how quantum mechanics can be understood as a space-time
theory provided that its spatial continuum is modelled by a
variable real number (qrumber) continuum. Such a continuum can be
constructed using only standard Hilbert space entities. The geometry of
atoms and subatomic objects differs from that of classical objects.
The systems that are non-local when measured in the classical
space-time continuum may be localized in the quantum continuum. We
compare this new description of space-time with the Bohmian picture of 
quantum mechanics.
\end{abstract}

\maketitle


\section{What is quantum space-time?}
Both modern mathematics and modern physics underwent serious foundational crises during the 20th century.

The crisis in mathematics occured at the beginning of the century and the main problem was
 to deal with certain infinities that are directly related to the concept of real number. 
 Poincar\'e \cite{poincare} explained this crisis in terms of different attitudes to infinity, related to Aristotle's
actual infinity and the potential infinity (the first attitude believes that the actual infinity
 exists, we begin with the collection in which we find the pre-existing objects,
  the second holds that a collection is formed by successively adding new members,
   it is infinite because we can see no reason why this process should stop). It led finally to the emergence of new, non-standard definitions of real numbers.

The crisis in physics concerns the interpretation of the quantum theory, the measurement problem and the question of non-locality.

In previous works we showed how in principle certain paradoxes of the quantum theory can be explained provided we enlarge our conception of number \cite{durt}.Our goal was to show how the basic axioms of quantum mechanics can be reformulated in terms of non-standard real numbers that we call qrumbers. It is our goal in the present paper to analyze non-locality and the concept of space-time at the light of the new conceptual tools that we developed in the past. 

Our main motivation is that most discussions of quantum mechanics use a background space-time that is 
the same as classical space-time, usually without any supportive arguments 
and even sometimes denying that quantum mechanics is a space-time theory. And yet many of the difficulties in understanding quantum phenomena derive from the use of classical space-time. We claim in the present paper that the space-time of quantum phenomena differs from that of classical phenomena in the nature of its continuum.  According to our theory \cite{durt}, the description of quantum phenomena requires a real number continuum that is not the classical continuum. It is not even a fixed element of the theory but varies with the quantum system in a way similar to the way the metric geometry of Einstein's general relativity varies with the physical system \cite{adelman2}. This is not 
part of the usual paradigm of quantum theory but adopting it enables us to reformulate the paradoxes of the standard interpretation when each quantum system has its own real number continuum.

The points of quantum space are identified with triplets of quantum real numbers, which we call qrumbers to help distinguish them from quantum numbers of standard quantum theory.  Qrumbers are real numbers that are taken as numerical values by quantum systems \cite{adelman2}. One important way in which they differ from standard real numbers is that each qrumber has a non-trivial extent to which it is valid. This extent depends upon the condition of the system. Moreover a copy of the standard reals is embedded in the qrumbers, in fact, to every extent, the standard rationals are dense in the qrumbers. 

In this approach, a non-standard qrumber is never obtained as the output from the measurement of a quantity because a measurement is a process in which a standard rational number is obtained as an approximation to the qrumber value of the quantity whose extent is conditioned by the measurement \cite{durt}. The empirical fact behind this is that measurements can only produce standard rational numbers.

If we accept the identification of classical points in standard Euclidean space with triplets of standard real numbers, then a single point of quantum space may be approximated by different classical points depending on how the measurements are made. For example, it is possible that a quantum particle localized in the vicinity of a point in quantum space is not localized in the neighbourhood of a single classical point. Furthermore a quantum particle with qrumber values for its position and momentum has a trajectory in qrumber space \cite{adelman2}. This does not contradict the Heisenberg uncertainty relations which only restrict the product of the ranges of standard real number values of positions and momenta that can be prepared or measured.\cite {durt}  The trajectories in qrumber space are obtained as solutions to equations of motion that are discussed in Ref.\cite{adelman2,durt}. They have the form reminiscent of the classical Hamiltonian equations of motion but instead they are formulated in terms of the qrumber values of position and momentum. In a sense this result evokes a picture reminiscent of the Bohmian picture in which trajectories in a generalised configuration space are associated with a quantum system. However the ontologies of the two pictures are different at the level of kinematics and dynamics.

In this paper the standard Hilbert space formulation of quantum mechanics 
is interpreted as a space-time theory by using an approach to the 
coordination problem similar to that suggested by Dieks \cite {dieks}. 
In his approach the numerical values of the position of a particle are taken 
to be attributes of the particle in the same way as its mass or charge. Then 
the assignment of numbers to the physical properties of a particle is made 
in such a way that the standard form of the physical laws governing the 
motion of the particle is maintained. We use a real number continuum given by
the sheaf of Dedekind reals $\RsubD(\EsubS)$ in the topos of sheaves on the   quantum state space $\EsubS$. A qrumber 
is a local section of the sheaf  $\RsubD(\EsubS)$ defined in section \ref{phasespace} \cite{maclane}, \cite{adelman2}.

Accordingly, the ontology of quantum particles is that of classical particles except that the numerical values of their attributes are given by qrumbers. 

This definition mixes standard concepts of the Hilbert space but also sophisticated concepts that were developed in the framework of non-standard analysis. Nevertheless it is sufficient for the comprehension of the nextcoming results to visualise qrumbers associated to a set of quantum observable as a cloud of standard real numbers that we obtain by computing the average values of these observables while the density matrix associated to the system varies in a certain open set or extent. In this view, sharp values for the observables that are characteristic of the classical picture (in terms of standard real numbers) gets replaced by ''fuzzy'' values.

Accordingly, the ontology of quantum particles is that of classical particles except that the numerical values of their attributes are given by qrumbers.

In this paper we focus on the changes in the concept of localisation. A particle that 
is localised in terms of the qrumber space may be non-localised when viewed 
from the classical real number space. For example, we consider Bell's experiment 
for two massive identical spin-1/2 particles \cite{bell}. When the standard description is 
expressed in terms of qrumbers we can show that the qrumber distance 
between the identical particles is zero at the time that the spin measurements 
in the classically separated Stern-Gerlach apparatuses are carried out.   
We compare this with the Bohmian picture \cite{bohm2} in which the separation is zero only in average \cite{durtbohm}. 

\section{What is a space-time theory?}

In this paper, we restrict our attention to space-time theories of particles the 
prototype of which was given by Newton in $\it{Principia}$ \cite {newton}. 
In it the motion of a material point particle is described by giving the values of 
its spatial coordinates as functions of time. The space consists of the set of values of the spatial coordinates available to the material particle. The space of a 
Newtonian particle is three-dimensional Euclidean space, that is, the spatial 
coordinates of a classical particle at a particular instant are given by a triplet 
of three standard real numbers. Each triplet is identified with a point in a three dimensional Euclidean geometry. The properties of the points are defined 
abstractly through Euclid's postulates which impose restrictions on the classes 
of real number continua that can be used to label the points, but the restrictions are 
not sufficient to uniquely define the real number continuum. Once a real number continuum is determined for the theory, then the numerical values of the spatial coordinates at an instant depend upon a frame of reference. The set of permissible frames of reference is invariant under the relativity group of the theory.

We define a space-time theory of a Galilean relativistic quantum system to be 
a theory of that system in which the particles have spatial coordinates that take 
numerical values at each instant of time. The set of values of the spatial 
coordinates available to the particles gives the spatial continuum, the set of 
instants available to the particles gives the time continuum. Mathematically
these identifications are only valid up to an isomorphism.

Newton in Ref.\cite{newton} introduced the concepts of absolute and relative space,

"Absolute space, in its own nature, without relation to anything
external, remains always similar and immovable. Relative space is some
movable dimension or measure of the absolute space, which our senses
determine by its position to bodies; and which is commonly taken for
immovable space."

We understand  this distinction in the following way.  When we have a physical 
theory which is expressed in mathematical terms we can understand absolute 
space as being just the abstract mathematical structure of space. For example, 
Newton's  absolute space is the purely axiomatic Euclidean geometry used in $\it{Principia}$.  Then the distinction between absolute and relative space
merely denotes the difference between an axiomatic geometry for space,
which is a purely abstract mathematical construction, and what Einstein
called the "practical" geometry of space that is obtained when the "empty 
conceptual schemata" of axiomatic geometry are coordinated with "real
objects of experience". The recognition and identification  of  the "real objects 
of experience" depends upon the structures that the theorist has imagined
previously. In 1930 Einstein \cite{einstein1} noted,

"It seems that the human mind has first to construct forms independently
before we can find them in things. Kepler's marvelous achievement is a
particularly fine example of the truth that knowledge cannot spring from
experience alone but only from the comparison of the inventions of the
intellect with observed fact." 
 
The purely abstract mathematical construction that lies behind the present 
approach is a special case of the idea of a topos first developed 
by Grothendieck, then by Lawvere and Tierney to replace set theory as 
the proper framework for mathematics \cite{maclane}. 
We will use a spatial topos, $\Shv(X)$, the category of sheaves 
on a topological space $X$. The standard real number 
continuum is the prototype of a class of real number continua, which we 
call the Dedekind reals $\RsubD(X)$, that are given by the sheaf of 
germs of continuous real valued functions on the topological space $X$. 
The open subsets of $X$ are the extents to which the numbers exist,
they give the truth values. Hence the internal logic is intuitionistic in 
general but is Boolean when $X$ is the one point space or when the 
topology on $X$ is trivial.

In the following paragraphs we assume that a physical attribute of a 
physical system is a quantity that would yield a single-valued, standard or classical real number if 
it were measured. Thus each of the three components of a position 
vector is a physical attribute.

In the standard formulations of Galilean relativistic quantum mechanics,  
in a given reference frame, at each instant of time, each physical attribute 
of a quantum system is represented by a self-adjoint operator $\hat A \in 
\mathcal A$, an  $O^{\ast}$-algebra \cite{inoue}. The states of the quantum 
system are normalized positive linear functionals on $\mathcal A$.  When 
restricted to the real algebra $\mathcal A_{sa}$ of self-adjoint operators 
the functionals map $\mathcal A_{sa}$ to $\RR$. The set of functionals
is called the quantum state space, $\EsubS$, of the system. The qrumbers of a system are the Dedekind real numbers $\RsubD(\EsubS)$ on its quantum state space.  

The classical mechanical analogue of quantum state 
space, $\EsubS$,  is isomorphic to the one point space $\{\ast\}$, so that 
the Dedekind real numbers for a classical system are $\RsubD(\{\ast\}) =  
\RR $, the standard real numbers. To see this, consider a system of massive 
particles in the Hamiltonian formalism of classical mechanics. In a given reference frame, at each instant of time, each physical attribute of the classical particles is represented by a real number, its value at that instant of time in 
that reference frame. Therefore the algebra $\mathcal A_{cl}$ of the
representatives of all physical attributes at one time in a given reference 
frame is isomorphic to the algebra of real numbers $\RR$. Then it follows 
that the set of normalized positive linear functionals on $\mathcal A_{cl}$ 
contains only the identity map. Therefore the space  of ``states''  for a 
classical mechanical system is the one point space and hence 
$\RsubD(\{\ast\}) =  \RR $.
 
The concept 'state'  employed in the preceding paragraph is different from that 
normally used in classical mechanics. Here, as in standard quantum mechanics, a state is a normalized positive linear functional on the algebra of physical 
quantities in a given reference frame at a given time. In classical mechanics, 
a state is given by a point $ ( q, p ) $ in phase space. The latter is both 
determinative, because $ ( q, p ) $ is the initial condition for Hamilton's equations 
of motion that uniquely determines the future states of the system, and generative, 
because every physical quantity is given by a real-valued function of  $ ( q, p ) $.  The quantum states do not have the generative property and therefore do not play the same role as the classical mechanical states. If the classical mechanical states are taken to be ontic, the quantum states can be taken to be epistemic.

\subsection{Arguments that QM is not a space-time theory}

 The following is a summary of the argument that
quantum mechanics is not a space-time theory \cite {dieks}.   

1. The Hilbert space formalism is self-sufficient and does not need a
space-time manifold as a background. There is no special role for
position, all physical quantities have the same status; they are all
represented by 'observables', i.e. by self-adjoint operators. 

2. Furthermore, the position and momentum quantities are represented by 
operators that do not commute. It then follows from the standard
interpretation that they cannot both have well-defined values at the same time. 
If the eigenvalue-eigenstate link is accepted (this is the rule which says that a quantity has a definite value only if its system is in an eigenstate of the corresponding observable) a particle cannot have both a well-defined
position and a well-defined momentum at any instant and hence cannot have
a trajectory because non-commuting operators don't have common eigenstates.  

3. The nub of the argument is that "the standard mathematical formalism of
quantum theory strongly deviates from its classical counterpart in that
physical magnitudes are not represented by functions on space (or on
phase space)". The operators used for the representation of physical
magnitudes in quantum theory are not $\it{automatically}$ associated with
real number values.

We think that these arguments do not prove that quantum
mechanics is not a space-time theory, rather they show that a
particular interpretation of the Hilbert space formalism of
quantum mechanics does not have a space-time picture.

1. The first point is dubious because many of the problems of the
standard interpretation of QM, such as the problems of locality, can be
related to the absence of quantum space. It has been asserted of this
interpretation that ``all the problems centre around the relation between -
on the one hand - the values of physical quantities, and - on the other
hand - the results of measurements''\cite{isham}. These values and results
are usually taken to be the same type of entities, by the precedent of
classical physics, but the ancient Greek ``measurement problem'' - that the
results of measurements should be rational numbers, but the value of the
ratio of the diagonal to the side of a square is not - should caution
us. The values of physical quantities can be a type of real number more
general than the numbers obtained as the results of measurements. 
Moreover the standard interpretation does not adequately distinguish the
representation of a physical quantity from the values that the physical
quantity may take. This appears to be a prejudice based on their
confluence in classical mechanics.

2.  The second point re-emphasizes the problems of the standard
interpretation. The eigenvalue - eigenstate link hypothesis is
ambiguous\footnote{This hypothesis is often
not listed in the axioms for QM, e.g. it is not in  Ref.\cite{cohen-tannoudji}, because 
operators don't always have eigenstates. In particular the position and momentum operators have no eigenvalues, although we can construct states that yield approximate eigenvalues \cite{vonneumann}.}; it is not clear whether it refers only to measured values.
Dirac states that ``a measurement always causes the system to jump
into an eigenstate of the dynamical variable that is being measured''\cite{dirac} 
but does not say from where the state has jumped, c.f., the
discussion in Bell \cite{bell}. 

We will re-interpret the standard mathematical formalism of QM to reveal an underlying space that generalises the space of classical mechanics. From this point of view, the main error of the standard interpretation was to not examine the
mathematical structure of quantum mechanics with sufficient care to
discover the spatial structure existing within it. It is worth noting that in the Bohmian approach \cite{bohm2} where the privileged observable (beable) is the position, the eigenvalue-eigenstate link hypothesis is not fulfilled in general, even when the observable that we consider is the position. Indeed if this hypothesis was fulfilled, the fact that trajectories are supposed to exist would imply that wave functions are not spread in space. The eigenvalue-eigenstate link is also not true when we consider velocities. This does not lead to any inconsistency in the Bohmian picture
 where instantaneous velocities are not directly measurable \cite{bohm2}.

3. Again this comment is true only if by "space" we mean the space of
\emph {standard} real numbers. It shows that quantum mechanics cannot
be represented as a space-time theory with standard real numbers. In
classical mechanics, the algebra of the physical quantities is
represented by real numbers that are identified with the entities that
represent their values. This leads to the identification of physical
quantities, such as the energy $H(p,q)$, that are given by functions on 
phase space with the values that their functions take at points on
phase space. It is this identification that is not maintained in quantum
mechanics, but we claim that classical space (or phase space) does not 
provide the underlying continuum for quantum mechanics. The qr-number 
values of physical quantities in quantum mechanics are given by continuous functions on the standard quantum state space (defined in section \ref{phasespace}).  

A brief review of the development of the real number concept is relevant,
especially in view of the identification of real numbers with points on
a straight line, initiated by Descartes, that opened the way to the use of
functions to describe the relations between geometrical objects and
points. The ancient Greeks, who developed what we call Euclidean geometry,
had not made that identification \cite{hartshorne}; magnitudes such as
lengths, angles, areas and volumes were measured using geometrical
motions of translations and rotations of line segments, triangles etc
aided by concepts of similarity and congruence that depended upon the
Archimedean principle that we discuss in section \ref{archimed}. It is not surprising that they only used ratios of natural numbers for if we only admit numbers that arise as the outputs from measurements we only admit rational numbers. 
It is not until we have a mathematical theory that describes what happens 
between the outcomes and inputs that we need numbers more general 
than rational numbers to ensure that the equations of the theory have 
numerical solutions. The properties of the equations and their solutions 
are used to deepen our understanding of the physical world.  It is in this 
sense that Dedekind's statement, ``Numbers are 
free creations of the human intellect, they serve as a means of grasping 
more easily and more sharply the diversity of things. ''\cite{dedekind},
may be understood.

\section{Qrumbers; Real numbers for quantum mechanics}
The numerical values taken by physical quantities in quantum mechanics
differ from standard real numbers in much the same way as variable functions differ from constants. Like functions they have domains of definition over which they may vary. Their definition comes from topos theory.
\begin{definition}
The qr-number continuum for a quantum particle is the continuum of Dedekind reals $\RsubD(\EsubS)$ given by the sheaf $\mathcal C(\EsubS)$ of continuous
real-valued functions in the topos $\Shv(\EsubS)$ of sheaves on
$\EsubS$.
\end{definition}
To apply this definition to quantum systems, 
we start from the standard Hilbert space formulation of quantum
mechanics as given by von Neumann \cite{vonneumann}. In it the  physical
quantities are represented by the self-adjoint operators defined on dense
subsets of Hilbert space or, in the case of bounded operators,
on the whole Hilbert space. These self-adjoint operators
belong to an $O^{*}$ algebra $\mathcal A$, called the algebra 
of observables, if the symmetric product is used they form a real 
algebra $\mathcal A_{sa}$ \cite{inoue}.
The states are the other main ingredient of the standard Hilbert space 
formulation of quantum mechanics. In the initial formulation the pure
states are given by elements (vectors) belonging to the Hilbert
space. The concept of state was later generalised to include impure
states which are given by positive bounded self-adjoint trace class
operators of trace 1. As noted above, the general concept of state is that of 
positive continuous linear functionals on $\mathcal A$ (see also section \ref{phasespace}). The pure
states are given by projection operators onto one dimensional subspaces
of the Hilbert space, which connects them to the original definition
in terms of Hilbert space vectors, at least up to a complex
multiplicative factor of modulus 1. We denote by $\EsubS$ the
Schwartz subspace of the state space described in the appendix.
 
In the standard theory $\Tr \hat \rho \Ahat$ is the average or expectation 
value of the quantity represented by $\Ahat$ in the state given by 
$\hat \rho$.  In our theory for each operator  $\Ahat \in \mathcal A_{sa}$ and 
each open subset $W \in \EsubS$, $\Tr \hat \rho \Ahat$ defines a real valued function with domain $W$. In the topos of sheaves on the topological space $\EsubS$, the sheaf of Dedekind reals $\RsubD(\EsubS)$ is isomorphic to the 
sheaf of continuous real-valued functions on $\EsubS$, \cite{maclane}, so that a 
qrumber defined to extent $W$ is a continuous function defined on $W$. 
With the appropriate topology on $\EsubS$ the functions defined using
$\Tr \hat \rho \Ahat$ are continuous and thus 
define qrumbers \cite{adelman2}.

It is worth noting that, beside the fact that it shows how non-standard numbers
 can be used in order to reformulate the quantum theory, 
our approach also makes it possible to reformulate the postulates of quantum mechanics
 in terms of average (expectation) values only. This must be put in parallel with the results 
 of S. Weigert \cite{weigert} who showed that one obtains, 
 using the expectation values 
of a quorum of quantities, a closed system of linear 
differential equations that describes the quantum evolution of a system of spin $s$. ¥

For a Galilean relativistic quantum mechanical system, $\EsubS$ is the set of 
all normalised, strongly positive linear functionals on the enveloping algebra
of the irreducible representation of the Lie algebra of the extended
Galilean group.  The irreducible representation of the enveloping Lie
algebra of the extended Galilean group, labelled by $ ( m, U, s)$ with
central element $\hbar I$, is unitarily equivalent to the tensor product
$M \otimes M_s$ of the $Schr\ddot odinger$ representation $M$ of the
algebra of the Canonical Commutation Relations  (CCR-algebra) generated
by the operators $ \{ \vec P,  \vec X , \hbar I \}$ with $M_s$ the
irreducible representation of dimension  $2s + 1$ of the Lie algebra of
the rotation group $SO(3)$ \cite{adelman2}. We choose the topology on $\EsubS$  to be the weakest that makes continuous all the functions of the form $a_Q( \hat \rho) = \Tr(\hat \rho  \hat A)  $ for self-adjoint operators 
$\hat A \in M \otimes M_s$. Then the functions $a_Q$ form a subobject $\AAA$
of $\RsubD(\EsubS)$ on $\EsubS$ \cite{adelman2}. Each qr-number defined to extent $W$ is either a continuous function of the $a_{Q}(W)$ or a constant real valued function on $W$.  The sheaf $\mathcal C(W)$ of continuous
real-valued functions over the open set $W \subset \EsubS$ can extended to a 
sheaf over $\EsubS$ by prolongation by zero \cite{swan}.

\section{Physical Interpretation \label{archimed}.}
A physical interpretation of these non-standard real numbers may be
linked, via the interpretation of a quantum state $\hat  \rho$ as
representing an ideal preparation process, to the  requirement that the whole
experimental arrangement must be included in the determination of
physical quantities \cite{bohr}. However our model differs from this
requirement in two important ways: firstly, the physical processes that
constitute the preparation process may occur naturally without the
intervention of experimentalists and, secondly, the preparation
procedures are represented by open sets of states. 

If a physical attribute of a particle is represented by the
self-adjoint operator $\hat A$ then the qrumber values that the
quantity can take are given by functions $a_Q(W)$ defined on open sets
$W$ by $a_Q(\hat \rho) = \Tr \hat \rho \hat A $ for all $\hat \rho \in W$.
The open set $W$ is the extent to which the quantum particle exists; we
say that $W$ is the ontic state of the particle. If the 
ontic state of the particle is $W$ then the qrumber value of the quantity 
represented by $\hat B$ is given by a function $b_Q(W)$ defined on $W$. 
In this sense the qrumber values of any quantity are determined by 
the ontic state $W$. An ontic state $W$ is specified by the set of all 
non-empty open subsets of $W$.
Since all conditions are defined pointwise, if $U \ne 
\emptyset$ and $ U \subseteq W$ then a qrumber $a_Q(U)$  
will satisfy any condition satisfied by $a_Q(W)$. For example, 
$\epsilon$ sharp collimation, see \cite{durt}, is a property of the ontic state $W$ 
because if it holds on $W$ it holds on all $V \subset W$. If $\vec \Xhat$ is the triplet of self-adjoint operators that represent the position of a particle then its qrumber positions are given by the continuous functions $\vec x_Q(W)$ on open sets $W$. The set of all its qrumber positions constitute the quantum particle's three dimensional space, whose geometric properties depend upon the structure of the underlying continuum of $\RsubD(\EsubS)$ \cite{stout}.
$\RsubD(\EsubS)$ is a field that is partially ordered but not totally
ordered, in particular, trichotomy does not always hold. 
Also it is not Archimedean in the sense that there are qrumbers
$a \in \RsubD(\EsubS)$ such that $a > 0$ but there is no natural number 
$n \in \mathbb{N}$ such that $n > a$. $\RsubD(\EsubS)$ is a complete
metric space with respect to a distance function derived from the norm
function $|\cdot|:\RsubD(\EsubS) --> \RsubD(\EsubS)$  that takes a to 
$max(a,-a)$ \cite{stout}. With this we can define a distance function between particle positions in $\RsubD(\EsubS)^{3}$ that is used to define localisation. 

Quantum localisation is different from classical localisation.  
For example, suppose that the z-coordinate of a particle has the qrumber 
value $z_Q(W)$, where $W$ is the union of disjoint open
subsets, $W =  U \cup V$ with $U \cap V = \emptyset$.
$z_Q(W)$ is a single qrumber value and hence represents a single
point on the z-axis in quantum space. But if there are standard real
numbers $z_1 < z_2 < z_3 < z_4$, with $z_1 < Z_{Q}(U) < z_2$,
$z_3 < Z_{Q}(V) < z_4$ and $z_2 \ll z_3$ then the single qrumber 
value may be viewed classically as a pair of separated intervals of
standard real numbers and hence non-localised. This example of the
difference between quantum and classical localisation can be used to
understand the 2-slit experiments. A particle may be localised
in terms of the qrumber values of its position but not localised in
terms of the classical standard real number values. The qrumber
distance between a pair of particles may be small even though the
classical distance between them is large.

In the Bohmian approach, non-locality \cite{epr,bohm2,bell} is related to the fact that
 the configuration space to which particles belong is not 3-dimensional but 3N dimensional. In our approach we generalise the concept of real number on which the concept of spatial point relies. 

\subsection{What is a quantum particle?}

A Galilean relativistic quantum particle is an entity localised in the qrumber
space with the following characteristics which generalize those of a 
classical particle \cite{bitbol}:

(i) it has permanent properties which always possess qrumber values that may be approximated by standard real number values which are revealed in observations expressed by counterfactual empirical propositions (such as 'if I had performed such and such experiments, I would have obtained such and such standard real number outcomes'),  

(ii) it has individuality,

(iii) it can be re-identified through time.

We must recognize the distinction between
intrinsic properties that are independent of the condition of the
particle, how it was prepared or what interactions it is undergoing and
the extrinsic properties that may depend upon the condition of the
particle. Particles with the same intrinsic properties are normally said
to be identical \cite{jauch}. For Galilean relativistic particles in
quantum mechanics, the intrinsic properties are the mass $m$, a positive
standard real, the internal energy $U$, a standard real, and the
spin $s$, a natural number or half a natural number. $( m, U, s)$ label
the irreducible projective unitary representations of the Galilean
group $\mathcal{G}$ \cite{levy}. The basic properties of the irreducible 
representations of $\mathcal{G}$ are given an appendix. 

It turns out that the
concepts of indistinguishability (:= two entities are indistinguishable
if they agree with respect to all their attributes) and identity (:= two
entities are identical if they are the same object)
are not equivalent in general \cite{dacosta}. 

Nevertheless, in quantum mechanics it is true that for all fundamental particles identity 
implies indistinguishability and vice versa. 
 
We will now study two particle systems in our approach.

\section{Two particle systems}

\subsection{Two different particles}

In the standard mathematical framework the Hilbert space for the quantum
mechanics of two different particles of unequal masses $m_1$ and
$m_2$, internal energies $U_1$ and $U_2$ and arbitrary spins $s_1$ and
$s_2$ is given by the tensor product
$\scrH(1) \otimes \scrH(2)$ of the Hilbert  spaces $\scrH(1) $ and $
\scrH(2)$ that are respectively the carrier spaces for the irreducible
projective unitary representations of the Galilean group $\mathcal{G}$
labelled by $(m_1,U_1, s_1)$ and $(m_2, U_2, s_2)$. For each
$j= 1,2$,  $\scrH(j) =  \scrH_j(ccr)\otimes \mathbb{C}^{2s_j +1}$
where $\scrH_j(ccr) = \mathcal{L}^{2}(\mathbb{R}^3)$ for both
$j = 1,2$.

We will write 
\begin{equation}
\scrH(1,2) = \scrH(1) \otimes \scrH(2).
\end{equation}

The physical quantities associated with particle 1 include 
operators, representing elements of the enveloping algebra of the Lie
algebra of $\mathcal{G}$, that are essentially self-adjoint on the
Schwartz subspace $\scrS$ of the spatial Hilbert space
$\mathcal{L}^{2}(\mathbb{R}^3)$ tensored with self-adjoint spin matrices
acting on $\mathbb{C}^{2s_1 +1}$.  To simplify the notation we will usually
denote a tensor product of operators associated with particle 1 by a
single symbol $A(1)$. Then the vector operators for particle $1$ are
position $\vec X(1) = \vec X \otimes I_{s_1}$, momentum  $\vec P(1) = \vec
P \otimes I_{s_1}$, angular momentum $\vec L(1) = \vec L \otimes I_{s_1}$
and spin
$\vec s(1) = I_{1} \otimes \vec s $, where $I_{s_1}$ is the identity
matrix on
$\mathbb{C}^{2s_1 +1}$ and $I_{1}$ is the identity operator on 
$\mathcal{L}^{2}(\mathbb{R}^3)$. We take the Schwartz space $\scrS(1)$ to
be the tensor product $\scrS \otimes \mathbb{C}^{2s_1 +1}$ . The physical
quantities associated with particle 2 have a similar tensor
product composition and will be simply denoted by operators $B(2)$ acting
on $\scrS(2) = \scrS \otimes \mathbb{C}^{2s_2
+1} \subset \mathcal{L}^{2}(\mathbb{R}^3) \otimes \mathbb{C}^{2s_2 +1}$. 
They include the vector  operators for position $\vec X(2) = \vec X
\otimes I_{s_2}$, momentum $\vec P(2) = \vec P \otimes I_{s_2}$, angular
momentum
$\vec L(2) = \vec L \otimes I_{s_2}$ and spin $\vec s(2) = I_{2} \otimes
\vec s $, where $I_{s_2}$ is the identity matrix on $\mathbb{C}^{2s_2 +1}$
and
$I_{2}$ is the identity operator on $\mathcal{L}^{2}(\mathbb{R}^3)$.
Because the tensor product $\scrS \otimes \mathbb{C}^{2s_j +1}$ is
denoted by
$\scrS(j)$ for $ j=1,2$, then $\scrS(1,2) = \scrS(1) \otimes
\scrS(2)$ represents the two particle Schwartz space. 
 
The algebra
$\mathcal{A}_{sa}(1,2)$ of operators representing physical quantities
associated with the compound system contains elements of the form
$A(1)\otimes B(2)$ as well as other operators, such the total energy
$H(1,2) = \Sigma_{j=1}^{3} P_{j}(1)^{2}/(2m_1)\otimes I(2) + I(1)\otimes
\Sigma_{j=1}^{3} P_{j}(2)^{2}/(2m_2) + V(1,2)$ where the interaction
operator $V(1,2)$  is generally not of the form $A(1)\otimes B(2)$.
When considered as part of the compound system the properties of
particle 1 are represented by operators of the form $A(1)\otimes I(2)$ and
the properties of particle 2 are represented by operators of the form
$I(1) \otimes B(2)$, with $I(j) = I_{j} \otimes I_{s_j}, j = 1, 2$.

Schwartzian state space $\EsubS(1,2)$ is the state space of the compound
system. $ \EsubS(1,2) $ is contained in the convex
hull of projections $\scrP(1,2)$ onto the one-dimensional subspaces of
$\mathcal{H}(1,2)$ spanned by unit vectors $\psi(1,2)$ belonging to the two particle Schwartz space $\scrS(1,2)$.
Each  state $\hat \rho\in\EsubS(1,2)$ is a trace class positive
bounded  operator on $\scrH(1,2)$ with trace $1$ which can be written as
$\hat \rho=\sum\lambda_{n}\scrP_{n}(1,2)$. For all
$n$, $\lambda_{n}\geq{0}$, $\sum\lambda_{n}=1$ and the $\scrP_{n}(1,2)$
are orthogonal projections onto one-dimensional subspaces in
$\scrS(1,2)$.  For each $A(1,2) \in \mathcal{A}_{sa}(1,2)$ we
define $ a(1,2)_Q(\hat \rho) = \Tr \hat \rho A(1,2)$ for $\hat \rho \in
\EsubS(1,2)$. $\EsubS(1,2)$ is given the weakest topology that makes all the  functions $a(1,2)_Q$ continuous.
 
Physical quantities that are represented by operators of the form
$A(1)\otimes B(2)$ have qrumber values $(a(1) \otimes
b(2))_{Q}(W)$ defined on open subsets $W \in \EsubS(1,2)$. Even though 
as operators  $A(1)\otimes B(2) = (A(1)\otimes I(2))(I(1)\otimes B(2))$ 
it is generally not true that $(a(1) \otimes b(2))_{Q}(W) = (a(1) \otimes
1(2))_{Q}(W))((1(1) \otimes b(2))_{Q}(W))$ because any open set $W$ will
contain non-product states from $\EsubS(1,2)$. Of course we can rewrite  
$(1(1) \otimes b(2))_{Q}(W) = b(2)_{Q}(W(2))$ where $W(2)$ is the open
subset of  $\EsubS(2)$ that is obtained by partial tracing $W \in
O(\EsubS(1,2))$ over an orthonormal basis in $\scrH(1)$. In a similar
fashion $(a(1) \otimes 1(2))_{Q})(W) = a(1)_{Q}(W(1))$ where $W(1)$ is the
open subset of 
$\EsubS(1)$ that is obtained by partial tracing $W \in O(\EsubS(1,2))$
over an orthonormal basis in $\scrH(2)$. Therefore $(a(1) \otimes
b(2))_{Q}(W) \neq a(1)_{Q}(W(1))\times b(2)_{Q}(W(2))$ unless for every 
$\hat \rho \in W, \hat \rho = \hat \rho(1) \otimes \hat \rho(2)$
where $\hat \rho(j) \in W(j) , j= 1,2$. Because quantum systems can be entangled, the latter condition is not satisfied in general and qrumbers do not always factorize.

Of particular interest are the qrumber values of the
kinematical variables of either of the two particles. For example, on an
open set 
$W \in \EsubS(1,2)$; the position vector  of particle 1 will have 
qrumber value $ \vec x(1)_{Q}(\tilde W(1))$,
the momentum vector of particle 1 will have qrumber value 
$ \vec p(1)_{Q}(\tilde W(1))$, while the 
position vector of particle 2 will have qrumber value  
$ \vec x(2)_{Q}(\tilde W(2))$ and the
momentum vector of particle 2 will have qrumber value  
$ \vec p(2)_{Q}(\tilde W(2))$, where for $j = 1, 2$, $\tilde W(j)$ is an open
subset of  $\EsubS(j)$ that is obtained by partial tracing $W(j)$ over an orthonormal basis in $\mathbb{C}^{2s_j +1}$, where $W(j)$ was obtained
by tracing  over an orthonormal basis in $ \scrH(k), k
\ne j$.  The qrumber distance between the two particles when
the compound system is in the open set $W \in \EsubS(1,2)$ is
$d(1,2)_{Q}(W) $ given by, where $\vec x(1) = \vec X(1) \otimes I_{s_j}$ etc.,
\begin{equation}
  \| (\vec x(1) \otimes I(2) - (I(1) \otimes \vec
x(2))_{Q}(W)) \|_{3} = \|\vec x(1)_{Q}(\tilde W(1)) -  \vec
x(2)_{Q}(\tilde W(2)) \|_{3} .
\end{equation}
That is, for each $\hat \rho(1, 2) \in W$,
\begin{equation}
 d(1,2)_{Q}(\hat \rho(1, 2)) = \|\Tr 
\hat \rho(1) \vec X(1) - \Tr \hat \rho(2) \vec X(2) \|_{3}.
\end{equation} 
where  $ \hat \rho(j)$ is obtained by partially tracing 
$\hat \rho(1,2)$ over 
an orthonormal basis in $\mathbb{C}^{2s_j +1}$ after having partially
traced over an orthonormal basis in $ \scrH(k), k \ne j$. $\|\vec v
\|_{3} = (\vec v \cdot \vec v)^{1/2}$ is the standard Euclidean norm of the
vector $ \vec v \in \RR^{3}$.

Consider the following situation; at $ t =  0$ particles 1 and 2 have the same
qr-number position and their qr-number momenta are of
the same magnitude but in opposite directions, that is,
$(\vec x(1) \otimes I(2))_Q(W) = (I(1) \otimes \vec x(2))_{Q}(W) $, and 
$(\vec p(1) \otimes I(2))_Q(W) = - (I(1) \otimes \vec p(2))_{Q}(W) $. If
the particles move freely then at any time $t > 0$, \cite{durt}

$(\vec x(1) \otimes I(2))_Q(W)(t) = (\vec x(1) \otimes I(2))_Q(W) +
{t\over m_1} (\vec p(1) \otimes I(2))_Q(W) $ and $(I(1) \otimes \vec
x(2))_{Q}(W)(t) =  (I(1) \otimes \vec x(2))_{Q}(W)  + {t\over m_2} (I(1)
\otimes
\vec p(2))_{Q}(W)$. Therefore at $ t > 0$ the qrumber
distance between the particles is 
\begin{equation}
 d(1,2)_{Q}(W)(t) = {(m_1 + m_2)\over (m_{1}m_{2})}t \| (\vec p(1) \otimes
I(2))_Q(W) \|_{3}.
\end{equation} 
where $\| (\vec p(1) \otimes I(2))_Q(W) \|_{3} = \| (I(1) \otimes
\vec p(2))_Q(W) \|_{3} $.

Each particle has its own continuous qr-number trajectory, viz., the
two particles move in opposite directions along a straight line in 
qr-number space.  But they do not have trajectories in standard real
number space.  However, at any instant of time, we can show that one of
the pair of particles is in a region of classical space while the other
is in a second classical region, well-separated from the first. This
is done by finding standard real numbers that approximate the
qrumbers $(\vec x(1) \otimes I(2))_Q(W)(t)$ and $(I(1) \otimes \vec
x(2))_{Q}(W)(t) $. That is, we can, in principle, determine an approximate
classical trajectory for each particle from its qrumber trajectory.This
contrasts with the standard interpretation in which Heisenberg's
uncertainty relations prevents us from determining a continuous trajectory
in classical space by which a particular particle could be labelled and
its path followed over time.

If the particles move freely then the total momentum is conserved so
that for all times $t$,
\begin{equation}
(\vec p(1) \otimes I(2))_Q(W)(t) = - (I(1) \otimes \vec p(2))_{Q}(W)(t).
\end{equation}
Thus each particle has both a definite qrumber value for its  position
and momentum for all times, these values remain correlated until some
interaction with the particles breaks the symmetry. Of course
measurement of the position and momentum of either particle will yield
approximate standard real number values which are in agreement with
Heisenberg's uncertainty relations \cite{durt}.

\subsection{Two identical particles}

When the two particles are identical then the standard quantum theory does not distinguish between them. 

In the standard mathematical framework the Hilbert spaces for the quantum
mechanics of two identical particles of mass $m \ne 0$, internal energy
$V$ and spin $s$ is given by the symmetric subspace (bosons), or
the anti-symmetric subspace (fermions), of the tensor product 
$\scrH(1,2) = \scrH(1) \otimes \scrH(2)$ of the Hilbert  spaces $\scrH(1)
$ and $ \scrH(2)$ that are identical copies of the carrier space for the
irreducible projective unitary representation $\mathcal U$ of the
Galilean group. $\mathcal U$  has labels $(m, V, s)$. For each
$j= 1,2$,  $\scrH(j) =  \scrH(ccr)\otimes \mathbb{C}^{2s +1}$
where $\scrH(ccr) = \mathcal{L}^{2}(\mathbb{R}^3)$. The symmetric
tensor product is written $(\scrH(1) \otimes \scrH(2))_{+}$ and the
antisymmetric tensor product is written $(\scrH(1) \otimes
\scrH(2))_{-}$.  We will write 
\begin{equation}
\scrH(1,2)_{\pm} = (\scrH(1) \otimes \scrH(2))_{\pm}.
\end{equation}
If $\Pi_{\pm}$ represent, respectively, the orthogonal projection
operators from $\scrH(1,2)$ to $\scrH(1,2)_{+}$ or to  $\scrH(1,2)_{-}$
then for any vector
$u \otimes v \in \scrH(1,2)$, $ \Pi_{\pm}(u \otimes v) = {1\over  2} (u
\otimes v \pm v \otimes u)$. It is easy to show that the operators
that represent physical quantities associated with systems of identical
particles must be invariant under permutation of the particles
 \cite{jauch}. For if we let $\mathcal P$ denote the unitary operator on
$\scrH(1,2)$ that interchanges the vector states of the two particles,
$\mathcal P (u \otimes v) = v \otimes u$, then
$\mathcal P = 2\Pi_{+} - I $ where $I$ is the identity operator on
$\scrH(1,2)$. We then observe that $\scrH(1,2)_{\pm}$ are the eigenspaces
of $\mathcal P$ corresponding  to the eigenvalues $\pm 1$.  An
operator respects the bosonic or fermionic identity of the particles if it
sends vectors in one of these eigenspaces to the same eigenspace,
therefore it must commute with $\mathcal P$. Hence the operators
that represent bosonic or fermionic physical quantities must be invariant
under permutations. 

That is the physical quantities associated with a system of two
identical particles, bosons or fermions, are represented by
operators
$A(1,2)$ that are symmetric under interchange of the labels of the particles.
The set of such operators form an algebra $(\mathcal{A}(1,2))_{+}$. 
Examples of operators $A(1,2) \in (\mathcal{A}(1,2))_{+}$ are $A(1)
\otimes A(2)$ and $A(1) \otimes I(2) + I(1) \otimes A(2)$ where the $A(j)$
are operators, built from products of operators representing elements of
the enveloping algebra of the Lie algebra of $\mathcal{G}$, that are
essentially self-adjoint on the Schwartz subspace $\scrS$ of the
configuration Hilbert space $\mathcal{L}^{2}(\mathbb{R}^3)$ tensored
with  symmetric spin matrices acting on $\mathbb{C}^{2s +1}$. The $ I(j) =
I_{j} \otimes I_{s_j}$ are the identity operators on 
$\scrH(j) =  \scrH(ccr)\otimes \mathbb{C}^{2s_{j} +1}, j = 1, 2$.

It follows that all states $\hat \rho(1,2)$ of a system of two identical
particles must also be symmetric under the interchange of the particles'
labels. This holds for both bosons and fermions.  Therefore state space
of a system of two identical particles is contained in the symmetric Schwarzian
state space $(\EsubS(1,2))_{+}$. The open subsets of $(\EsubS(1,2))_{+}$
are the restrictions of the open subsets of $\EsubS(1,2)$ to
$(\EsubS(1,2))_{+}$. They give the weakest topology that makes
continuous all the  functions $a(1,2)_{Q}$ from $(\EsubS(1,2))_{+}$ to
$\mathbb{R}$;
\begin{equation}a(1,2)_{Q}(\hat \rho(1,2)) = \Tr
\hat \rho(1,2) A(1,2)\end{equation} for 
$A(1,2) \in (\mathcal{A}(1,2))_{+}$, when
$\hat \rho(1,2)
\in (\EsubS(1,2))_{+}$. In particular, these functions are
continuous with respect to the topology on $(\EsubS(1,2))_{+}$  given by
the  restriction of the trace norm topology
$\nu(\hat \rho(1,2)) = \Tr |\hat \rho(1,2)|$ to $(\EsubS(1,2))_{+}$.

The kinematical quantities associated with a system of two identical
Galilean particles of mass $m \ne 0$ and spin $s$ have qrumber values 
defined to extents $W \in O((\EsubS(1,2))_{+})$ by
continuous functions; 
$\vec x (1,2)_{Q}$,$\vec p (1,2)_{Q}$,$\vec L (1,2)_{Q}$, defined for 
$\hat \rho (1,2) \in W$
by $\vec x (1,2)_{Q}(\hat \rho (1,2)) = \Tr \hat \rho (1,2) \vec X(1,2)$, 
where  $\vec X(1,2) = \vec X(1) \otimes I(2) + I(1) \otimes \vec X(2) $,
etc. The two particle system has a trajectory in qrumber space that is defined by
the time evolution of these values \cite{durt}. To obtain trajectories for the
individual particles inside this system we define their kinematical
quantities following the method used when the particles were not
identical.    

\begin{proposition}
In a system of two identical particles, the qrumber value of
any property of particle 1 is the same as the qrumber value of the corresponding
property of particle 2 when both qrumbers are defined on the
same open subset $W(1,2) \subset (\EsubS(1,2))_{+}$.
\end{proposition}
\begin{proof}
Suppose a physical property of a single particle of type $(m,V,s)$ is
represented by the operator $A$, so that when considered as part of the
compound system the property of particle 1 is represented by an operator
of the form $A(1)\otimes I(2)$ and the property of particle 2 is
represented by an operator of the form $I(1) \otimes A(2)$. Either 
$ A(1) = A_1 \otimes I_{s_1}, A(2) = A_2 \otimes I_{s_2} $ when $A$
represents a kinematical quantity or $ A(1) = I_{1} \otimes A_{s_1}, A(2) =
I_{2} \otimes A_{s_2}$ , when A represents a spin variable.

Consider the operator $B(1,2) = A(1)\otimes I(2) - I(1) \otimes A(2)$.
$B(1,2) = - B(2,1)$ is skew-symmetric with respect to the interchange of
particle labels. Therefore for any state $\hat \rho(1,2) \in
(\EsubS(1,2))_{+}$,
\begin{equation}
\Tr \hat \rho(1,2)B(1,2) =  - \Tr \hat \rho(1,2)B(1,2)
\end{equation}
because $ \Tr \mathcal P^{-1} \hat \rho(1,2) \mathcal P \mathcal P^{-1} B(1,2)
\mathcal P = \Tr \hat \rho (2,1) B(2,1) $.
Therefore 
\begin{equation}
\Tr \hat \rho(1,2) (A(1) \otimes I(2)) = \Tr \hat \rho(1,2) (I(1) \otimes A(2))
\end{equation}
\end{proof}

It is important to be clear that this result says only that when
particles 1 and 2 are identical then the qrumber values of the
physical quantities of particle 1 are the same as those of particle 2.
This does not imply that the measured values of quantities associated
with particle 1 must be the same as the measured values of the quantities
associated with particle 2. Nevertheless this result has considerable
consequences for the concept of locality for identical particles, because
it says that for each $\hat \rho(1, 2) \in W(1,2)$,
\begin{equation}
 \Tr \hat \rho(1, 2) (\vec X(1) \otimes I_{s_1})
\otimes I(2) = \Tr \hat \rho(1,2) (I(1) \otimes \vec X(2) \otimes I_{s_2}).
\end{equation}
Therefore the qrumber distance,  $d(1,2)_{Q}(W(1, 2)) = \|
(\vec X(1))_{Q}(W(1,2)) - ( \vec X(2))_{Q}(W(1,2)) \|_{3}$, between identical particles is
zero.
\subsubsection{EPR-type experiment} Let us now consider an experiment which uses two identical massive particles that are produced in such a way that the sum of their momenta is zero. The experimenter prepares the momenta to be along the line of a standard real number vector  $\vec a$. The two particle system is defined to an extent that is given by an $\epsilon$-neighbourhood, $W(1,2)$, of the symmetrised pure state $\Pi_{\pm}$ that projects onto the vector ${1\over  2} (u_L(1) \otimes v_R(2) \pm v_R(1) \otimes u_L(2))$ where $L$ and $R$ refer to wave packets propagating along opposite directions following the line parallel to $\vec a$.

If $\epsilon $ is small enough, the reduced states of the single particles belong to open sets that are a fifty-fifty convex combination of an $\epsilon$-neighbourhood of the projector onto $u_L$ and an $\epsilon$-neighbourhood of the projector onto $v_R$ (see section \ref{phasespace}). The qrumber value of the position and velocity of an individual particle are then close to zero for all times for each particle (both 1 and 2). We physically interpret this property as follows: as a consequence of indistinguishability, the individual particle trajectories reduce to those of the centre of mass of the system. If now we consider two particles trajectories in the qrumber approach, this is no longer true. Indeed, in our approach, we can describe the position of the pair by the qrumbers $ {1\over 2} ( \hat x^i(1)) \otimes  \hat x^j(2))_{Q}(W(1,2))$, ($i,j=1,2,3$). If we compute their value, up to negligible corrections that are proportional to $\epsilon$, using the appropriate system of reference axes we obtain the value zero for all except  
 $  ((\vec a.\vec x)(1) \otimes (\vec a.\vec  x)(2))_{Q}(W(1,2))$ which has a qrumber value on $W(1,2)$ which is close to the standard real value $-(v.t)^2$.  $v$ is the absolute value of the velocity of each component along the $\vec a$ direction and $t$ the time that has elapsed since their emission from the source.

 Combining this information about the trajectories with our previous observation of the fact that the center of mass does not depart from the origin, we can infer the correct picture for the pair trajectory, i.e. that one particle moves to the left and the other moves to the right, with equal speed $v$.
 
 It is interesting to note that, similar to what happens in quantum optics, the pictures that we get about one and two particle trajectories are, in a sense, complementary and, to some extent,  independent. This is not true in the Bohmian approach where individual trajectories are supposed to contain all the relevant information about the physical reality of quantum systems. Actually, in the exemple that we considered above, the two particles are entangled which explains why the knowledge of the pair does not reduce to the knowledge of its single components. The explanation of non-local features of the system is also different in the Bohmian approach and in ours because in the former it is due to the non-locality of the quantum potential and undirectly to the existence of an absolute spatio-temporal reference frame while in the latter non-locality is a property of conventional space-time, not of the quantum space. Moreover the measurement process itself is non-local as we shall now discuss with the help of the example provided by the EPR-Bohm-Bell experiment that we shall discuss now.
 
\subsubsection{EPR-Bohm-Bell  experiment for identical particles}

Consider an EPR-Bohm-Bell  experiment which uses
two identical spin one-half massive particles. 
We can obtain the usual quantum 
mechanical results for the experiment while maintaining that the two 
identical particles always have a qrumber position and hence are localised 
in qrumber space.

 The two particle system is now defined to an extent that consists of an $\epsilon$-neighbourhood, $W(1,2; s_1,s_2)$ of the symmetrised pure state $\Pi_{\pm}\otimes \pi(s_1,s_2)$ that projects onto the vector $ {1\over  2} (u_L(1)\otimes v_R(2) \pm v_R(1) \otimes u_L(2))\otimes( \vert +_{s_1}> \otimes \vert -_{s_2}>+\vert -_{s_1}> \otimes \vert +_{s_2}>)$ where $L$ and $R$ label the opposite directions along the line parallel to $\vec a$ and $ \vert \pm_{s_j} >$ represents a spin up (down) polarisation state along the directions $\vec b$ orthogonal to $\vec a$ for particle $j$. If we trace over the spin degrees of freedom the pure state $\Pi_{\pm}\otimes \pi(s_1,s_2)$ reduces to the pure spatial state $\Pi_{\pm}$ which is of the same form as that used for spinless particles. Therefore before the spins interact with the magnetic fields, the discussion of the previous section applies; the individual particle trajectories reduce to those of the centre of mass of the system and the qrumber values  of position and velocity of the individual particles are then close to zero for all times for each particle. It is only when the two particle trajectories are calculated that we see that one particle travels to the left and the other to the right, but we cannot tell which.

 In order to carry out an EPR-Bohm-Bell  experiment we must check the spin correlations of the particles. This can  be done by letting them pass through Stern-Gerlach devices with magnetic fields orthogonal to the direction $\vec a$. The positions of the particles are then measured after their passage through the Stern-Gerlach devices.

As we discuss with greater detail in Ref.\cite{durt} there exist several possible ways to describe the evolution of the system in the absence of measurement. For instance one could let it evolve according to the standard unitary (Schroedinger) evolution, or one could define an Hamilton-like evolution at the level of the qrumbers. In the present case, taking account of the gyromagnetic coupling of the spins, the two evolutions would be 
equivalent up to $\epsilon$ when the system is defined to an extent that consists of an $\epsilon$ neighbourhood $W(1,2; s_1,s_2)$ around the symmetrised pure state $\Pi_{\pm}\otimes \pi(s_1,s_2)$. 

 By choosing $\epsilon$ small enough the trajectories will be enclosed in regions of volume comparable to that due to the spread of the wave function. In well conceived experiments
 we measure the positions of the particles with detectors larger than this spread \cite{durtthesis}, so that it does not really matter which type of evolution we adopt in our description of the evolution of the system, provided that the time of passage through the magnetic region of the Stern-Gerlach devices is short enough.
 
After the passage through these regions and before reaching the detectors, the two particle system is defined to an extent given by an $\epsilon$-neighbourhood of the projection $\Pi_{\pm}^{s_1,s_2}(1,2)$ onto the vector 
$\Psi_{\pm}^{s_1,s_2}(1,2)$

 $= {1\over 2} (u^+_L(1) \otimes v^-_R(2)\otimes \vert +s_{L}(1)> \otimes \vert -s_{R}(2)> (\pm ) \  v^-_R(1) \otimes u^+_L(2))\otimes \vert -s_{R}(1)> \otimes \vert +s_{L}(2)> +  u^-_L(1) \otimes v^+_R(2)\otimes \vert -s_{L}(1)> \otimes \vert +s_{R}(2)> (\pm)  \  v^+_R(1) \otimes u^-_L(2))\otimes \vert +s_{R}(1)> \otimes \vert -s_{L}(2)>)$
  where $L^{\pm}$ and $R^{\pm}$ refer to wave packets propagating to the upper (lower) left and right parts of the plane spanned by $\vec a$ and $\vec b_{K}, K = L, R$, the direction of the magnetic field on the left or right. $\pm s_{K}(j)$ denotes spin up/down in the direction of $\vec b_{K}, K = L, R$ for the $jth$ particle. 
  
Tracing over the spin variables we obtain a reduced spatial state $P_{\pm}(1,2)$ given by ${1\over 4}(
P_{u_L^+(1)} \otimes P_{v_R^-(2)}  + P_{u_L^-(1)} \otimes P_{v_R^+(2)} + P_{v_R^+(1)} \otimes P_{u_L^-(2)}  + P_{v_R^-(1)} \otimes P_{u_L^+(2)} )$. If the wave packets are assumed to be non-overlapping narrow Gaussians then the single particle reduced state for the $jth$ particle is $\rho_{\pm}(j) =
{1\over 4}(  P_{u_L^+(j)} + P_{u_L^-(j)} +  P_{v_R^+(j)} + P_{v_R^-(j)})$, which is the same state for both bosons and fermions so we'll suppress the $\pm$.The
extent that the $jth$ particle exists is given by an $\epsilon$-neighbourhood of $\rho(j)$.  

If $\epsilon $ is small enough, the $\epsilon$-neighbourhood of $\rho(j)$ is a convex combination with equal weights of $\epsilon$-neighbourhoods of the pure state projectors onto the four vectors $u^{\pm}_L,  v^{\pm}_R$. 

  In each reduced spatial state $P_{\pm}(1,2)$, the pure states for particles 1 and 2 are paired. If we associate a segment of a qrumber trajectory with a single particle pure state via its $\epsilon$-neighbourhood then the qrumber trajectory for the identical particles is composed of a pair of distinct trajectories in the crossing diagram $(>--<)$. In the first of the pair, one of the particles, we cannot tell which, is at the top of the descending branch of the cross on the left and the other particle is at its bottom on the right; in the second of the pair, one of the particles, we cannot tell which, is at the top of the ascending branch of the cross on the right and the other particle is at its bottom on the left.

This situation clearly differs from the Bohmian picture in which either the ascending branch is selected or the descending one, but not both at the same time \cite{durtbohm}.
 
As we discuss in  Ref.\cite{durt}, it is at the moment of the measurement, when detectors placed beyong the region of the magnetic field click, one of the two branches of the qrumber trajectory is selected in a process equivalent to the collapse of a wave-function. We show in that paper that the probability associated to the respective branches necessarily obeys the Born rule in order to derive a self-consistent formulation of the qrumber approach. Clearly the interaction with the measuring apparatus is non-local in the usual sense when the  branches of the cross on the left and right are separated by spacelike distances (where we refer here to distances measured in standard Euclidean space).

In a sense the problem of non-locality for identical particles can be seen
to disappear when qrumbers are used because the qr-number distance
between the particles 1 and 2 is zero, provided that we accept that space-time of
quantum systems is described by its qrumber continuum rather than the classical real number continuum.
 
\section{Conclusions}
We have shown that if we are willing to accept that the spatial continuum is 
not given a priori but is an artefact of the theory then Galilean relativistic quantum mechanics has a space-time that is as real as the standard space-time of classical 
physics. Galilean relativistic atoms and sub-atomic particles exist in a space
whose points are labelled by qrumbers rather than standard real numbers, they 
move along trajectories in the space of qrumbers. Their properties have
qrumber values always. In this setting the ontology of Galilean 
relativistic atoms and sub-atomic particles is similar to that of classical 
particles except that the values taken by their attributes are qrumbers 
not standard real numbers. Accepting that we can have a non-standard spatial continuum for a physical theory allows us to better understand the theory, for 
example the problems of non-locality in the standard theory of quantum mechanics 
disappear when the qrumber spatial distance between quantum particles is used.
For example, we showed in an explicit example of a Bell-type experiment how identical particles have zero qrumber spatial separation and hence their
interactions are in a sense not non-local. The mystery of the two slit experiment is removed 
because the qrumber position of a quantum particle can be such that the one qrumber position can cover more than one standard spatial position of a classical 
particle.

Nevertheless, the quantum wholeness remains present because, due to the presence of entanglement, 2 particle behavior does not reduce to 1 particle behavior.

That the quantum space of a quantum system is different from standard Euclidean 
space does not reduce the efficacy of the space-time view for understanding physics.
The standard real number continuum of classical physics presents much the same class of difficulties as the qrumber continuum: we never can observe all the points on the trajectory of a moving particle, there are many more points in the continuum than we can ever observe. There is one important difference, the points of quantum space are faithfully represented by qrumbers which exist to varying extents reflecting the conditions on the physical system under which the quantities have a value. The dependence of the condition of the physical system raises new possibilities. The 
universality of the numerical values taken by quantities is an idealisation that has
real limitations; even in classical physics the length of a metal rod depends upon 
the ambient temperature. 
This paper only discusses Galilean relativistic particle theories. we think that will be possible to extend this approach to different relativistic quantum field theories when 
their symmetry groups are used in place of the Galilean group.

\section{Appendices.}
\subsection{The standard mathematical formalism}
We review the mathematical formalism of standard quantum
mechanics.

A Galilean relativistic particle of mass $m$, internal energy $U$ and
spin $s$ is associated with an irreducible projective unitary
representation of the Galilean group $\mathcal{G}$ \cite{levy}.
$\mathcal{G}$ is parameterised as follows  
$ g = (b, \vec a, \vec v, R)$:
$b\in \RR$ for time translations, $\vec a \in \RR^{3}$ for space
translations, $\vec v \in \RR^{3}$ for pure Galilean transformations or 
velocity translations, $R \in SO(3)$ for rotations about a point.
Classically  $g \in \mathcal{G}$ acts upon the classical space-time
coordinates sending $(\vec x, t)$ to $(\vec x', t') = (R\vec x + \vec a +
t\vec v , t + b)$ then 
\begin{equation}
 (b', \vec a', \vec v', R') (b,\vec a, \vec v, R)
= (b' + b, \vec a' + R'\vec a + b\vec v', \vec v' + R'\vec v, R'R)
\end{equation}

The irreducible projective unitary representation $g \in \mathcal{G}
\mapsto \mathcal{U}(g)$, labelled by $ ( m, U, s)$ where $m$ 
is a positive real number, $U$ is a standard real number and $s$ is a
natural number or half a natural number, acts on the Hilbert
space $\scrH :=
\mathcal{L}^{2}(\mathbb{R}^3) \otimes \mathbb{C}^{2s +1}$. The elements of
$\scrH$ are $(2s + 1)$-component vectors of square integrable functions
$\{ \psi_{i} (\vec x): \vec x \in \RR^{3} \}_{i = -s}^{s} $. The
corresponding space-time functions $\psi_{i} (\vec x, t)$ are defined
using the generator $H = {1\over 2m} \vec P\cdot \vec P + U$ of the time
translations,
$\psi_{i} (\vec x, t) := (exp(-i Ht/\hbar) \psi_{i}) (\vec x)$.  Then

\begin{equation}\mathcal{U}(g)\psi_{i} (\vec x, t) =
 exp(i
\alpha_{m} ) \Sigma_{j} D_{ij}^{s}(R)
\psi_{j}(R^{-1}( \vec x - \vec v(t - b) - \vec a), t - b)
\end{equation}
where $\alpha_{m} = \alpha_{m}(b,\vec a, \vec v, R ; \vec x, t) :=
[-{1\over 2}m\vec v\cdot \vec v (t - b) + m\vec v\cdot (\vec x - \vec a
)]/\hbar$ and
$D_{ij}^{s}(R)$ are the matrix elements of the irreducible projective
representation of $SO(3)$ on $\mathbb{C}^{2s +1}$.

If $H$ is the infinitesimal generator of the subgroup of time
translations, $P_1, P_2, P_3 $ are the infinitesimal generators of the
subgroup of spatial translations along the three orthogonal axes of the
standard basis, $K_1, K_2, K_3 $ are the infinitesimal generators of the
subgroup of velocity translations along those axes and
$J_1, J_2, J_3 $ are the infinitesimal generators of the
subgroup of rotations around those axes then 
the Lie algebra of the extended Galilean group, with elements 
$(\theta, g); \theta \in \RR, g \in \mathcal{G}$, is generated by $ H,
\vec P, \vec K, \vec J $ and the central element $\hbar I$ whose Lie
brackets with all the other 10 elements vanish. The other Lie brackets
are 
\begin{equation}
[J_i,A_j] = \epsilon_{ijk}A_k,  \vec A = \vec P, \vec K, \vec J;
\smallskip [H, B_i] = 0, \vec B = \vec P, \vec J; 
\smallskip [H, K_i] = - P_i.
\end{equation} and
\begin{equation}
[K_i,K_j] = 0, [P_i,P_j] = 0, [K_i,P_j] = m \hbar I \delta_{ij}.
\end{equation}
As well as $\hbar I$, the element $ U:= H - {1\over 2m} \vec P\cdot \vec
P$ of the enveloping algebra commutes with all the infinitesimal
generators.
$U$ may be identified as the internal energy of the particle. We usually
take $U = 0$. The vector position operator for the particle is $ \vec X =
{1\over m} \vec K$, the  vector operator $ \vec L = \vec X \times \vec P $
is the orbital angular momentum. The spin vector operator is $\vec S =
\vec J - \vec L$. The operator $\vec S \cdot \vec S$ commutes with all the
elements of the Lie algebra  and in any irreducible unitary
representations $\vec S \cdot \vec S = s(s + 1)$ with $s$ an integer or
half-integer.

The irreducible representation of the enveloping Lie algebra of the
extended Galilean group labelled by $ ( m, U, s)$ with central element
$\hbar I$ is unitarily equivalent to the tensor product $M \otimes M_s $
of the $Schr\ddot odinger$ representation $M$ of the algebra of the
Canonical Commutation Relations  (CCR-algebra) generated by the operators
$ \{ \vec P,  \vec X , \hbar I \}$ with the irreducible matrix
representation $M_s$, of dimension $2s + 1$, of the Lie algebra of the
rotation group $SO(3)$.  

The $Schr\ddot odinger$ representation of the CCR-algebra $M$ is the
representation in which the Hilbert space is  $\scrH(ccr) =
\mathcal{L}^{2}(\mathbb{R}^3)$.
$X_{j}$ is represented by multiplication by the real variable $x_{j}$
and $P_{j}$ by $(1/i)$ times the operator of differentiation with
respect to $x_{j}$.  Let $\scrS({\RR^{3}})$ denote the Schwartz
space of infinitely  differentiable functions of rapid
decrease on $\RR^{3}$. Then the physical quantities are
represented by self-adjoint  elements in the closure $\Mbar$
of the CCR-algebra $M $, where $\bar M$ is the smallest closed
extension of $M $, 
$\Mbar = \bigl\{\tilde X\bigm| X \in M\;\allowbreak
\text{and}\allowbreak
\;\tilde X\allowbreak
\text{is the restriction  to }\scrS({\RR^{3}}) 
\text{ of the Hilbert space closure of}\allowbreak
\;X\bigr\}$.  Following the definitions of Powers \cite{inoue}, $M$ is
essentially  self-adjoint because the adjoint $M^{*}$ of $M$ equals the 
closure $\bar M$ of $M $. 

The irreducible representations of the Lie algebra of $SO(3)$ are
labelled by the half integer s, $s = 0,{1\over 2}, 1,{3\over 2}, 2,...$,
and the dimension of the representation is $(2s + 1)$. The elements of
the Lie algebra are represented by self-adjoint matrices acting on
$\mathbb{C}^{2s +1}$, they form the matrix algebra $M_s$.

The physical quantities associated with a particle are represented by 
operators $\hat A$ that are the tensor product of operators, $\hat A =
\hat C \otimes \hat S$. $\hat C$ represents an element of the enveloping
algebra of the Lie algebra of
$\mathcal{G}$ and is essentially self-adjoint on the Schwartz  subspace
$\scrS({\RR^{3}})$ of the Hilbert space
$\mathcal{L}^{2}(\mathbb{R}^3)$. 
$\hat S$ is a self-adjoint spin matrix acting on $\mathbb{C}^{2s + 1}$.

\subsection{Qrumbers\label{phasespace}}
We use a real number continuum given by
the sheaf of Dedekind reals $\RsubD(\EsubS)$ in the topos of sheaves on the   quantum state space $\EsubS$. A qrumber 
is a local section of the sheaf  $\RsubD(\EsubS)$ \cite{maclane}, \cite{adelman2} 
$\EsubS$ is  the Schwartz subspace of the state space, consisting of those states $\hat \rho$ that are rapidly decreasing convex combinations of  projection operators
$\hat P_{j}$ onto one dimensional subspaces spanned by vectors in Schwartz
space $\scrS$, that is,  $ \hat \rho = \Sigma_{j=1}^{\infty} \lambda_{j} \hat
P_{j}$ with $ \lambda_j \geq 0 $ and $ \lim_{j\to \infty} \lambda_{j} j^{n} = 0$
for all positive integers $n$ \cite{adelman2}.

The topology on $\EsubS$  is the weakest that makes continuous all the functions of the form $a_Q( \hat \rho) = \Tr(\hat \rho \cdot \hat A)  $ for self-adjoint operators 
$\hat A \in M \otimes M_s$. Then the functions $a_Q$ form a subobject $\AAA$
of $\RsubD(\EsubS)$ which is the sheaf of locally linear functions
on $\EsubS$ \cite{adelman2}. Each qr-number defined to extent $W$ is either a continuous function of the $a_{Q}(W)$ or a constant real valued function on $W$.  The sheaf $\mathcal C(W)$ of continuous
real-valued functions over the open set $W \subset \EsubS$ can extended to a 
sheaf over $\EsubS$ by prolongation by zero \cite{swan}.

One useful type of open set is the $\epsilon$-neighbourhood of a state $\rho_{0}$.
\begin{definition}
The $\epsilon$-neighbourhood of a state $\rho_{0}$ is $\nu (\rho_{0}; \epsilon) = 
\{ \rho : Tr \vert \rho - \rho_{0} \vert < \epsilon \}$. That is, it is an open ball, in the trace norm topology, of radius $\epsilon$ centred on $\rho_{0} $.
\end{definition} When $\epsilon$ is small, we can get a good idea of the properties of $a_Q(\nu (\rho_{0}; \epsilon))$ from those of $Tr ( \rho_{0} \cdot \hat A)$.
 
The following result is used in the discussions of identical particles.
Let $\Psi (1,2) = {1\over \sqrt 2}( \psi_{+}(1) \psi_{-}(2) \pm \psi_{-}(1) \psi_{+}(2) )$ be an entangled wave function of particles 1 and 2, the normalised single particle wave functions being orthogonal, 
$< \psi_{+}(j),  \psi_{-}(j)> = 0, j = 1, 2$. 
The corresponding pure state is $P_{\Psi(1, 2)} = {1\over 2} 
( P_{\psi_{+}(1)} \otimes P_{\psi_{-}(2)} + P_{\psi_{-}(1)} \otimes P_{\psi_{+}(2)} \pm 
V_{+, -}(1) \otimes V_{ -, +}(2) \pm V_{-, +}(1) \otimes V_{ +, -}(2))$, where the partial isometries $V_{ +, -}(j) = \vert \psi_{+}(j)> < \psi_{-}(j) \vert  = V_{ -, +}(j)^{\ast}, j = 1, 2$. The reduced state of particle j is the mixed state, 
$\rho_{0}(j) = {1 \over 2} ( P_{\psi_{+}(j)}  +  P_{\psi_{-}(j)} )$, for both bosons and fermions.

\begin{lemma}
If $\epsilon < 1$ then the $\epsilon$-neighbourhood of $\rho_{0}(j)$ does not contain any pure state.
If $\epsilon$  is small enough it essentially only contains fifty-fifty convex combination of the  $\epsilon$-neighbourhoods of $P_{\psi_{+}(j)} $ and
$P_{\psi_{-}(j)} $, that is, $\nu(\rho_{0}(j) ;  \epsilon) = {1\over 2} \nu(P_{\psi_{+}(j)} ;  \epsilon)  +
{1\over 2} \nu(P_{\psi_{-}(j)} ;  \epsilon), j = 1, 2 $.
\end{lemma}
\begin{proof}
We will prove the result for particle 1.

If $\chi (1)$ is a unit vector with $< \chi(1) \vert \psi_{\pm}(1) > = \gamma_{\pm}$ then the vectors $\{ \psi_{+}(1),\psi_{-}(1), \chi(1) \}$ span a subspace of dimension at most 3. Then it is easy to calculate $Tr \vert  P_{\chi(1)} - \rho_{0}(1) \vert = 1 + ( 1 - |\gamma_{+}|^2 - |\gamma_{-}| ^2) ^{{1\over 2}}$ which is less than $\epsilon$ only if $\epsilon > 1$. Therefore, when $\epsilon < 1$  every state $\rho  \in  \nu (\rho_{0}(1) ;  \epsilon)$ is  impure. 

Now ${1\over 2} \nu(P_{\psi_{+}(1)} ;  \epsilon)  + {1\over 2} \nu(P_{\psi_{-}(1)} ;  \epsilon)$
is contained in $\nu(\rho_{0}(1) ;  \epsilon)$. Because if $\rho = {1\over 2}\rho_{+} + {1 \over 2}\rho_{-}$
with $\rho_{j} \in \nu (P_{\psi_{j}(1)} ; \epsilon)$ for $j = +, -$, then since $\nu $ determines a norm on the space of trace class operators,
\begin{equation} 
Tr \vert \rho - \rho_{0}(1) \vert \leq { 1 \over 2}(Tr \vert \rho_{+} - P_{\psi_{+}(1)} \vert + Tr \vert \rho_{-} - P_{\psi_{-}(1)} \vert) < \epsilon. 
\end{equation}
There are other states in the $\epsilon$-neighbourhood of $\rho_{0}(1)$. They are of the form, $\rho = \lambda P_{\psi_{+}(1)} +  (1 - \lambda)P_{\psi_{-}(1)} $  where  ${1\over 2} - {\epsilon \over 2} <  \lambda <  {1\over 2} + {\epsilon \over 2}$. They are impure states
which are convex combinations of the same two states as $\rho_{0}(1)$ is. When $\epsilon$
is small they behave like  $\rho_{0}(1)$ itself.
\end{proof}

Therefore if a system of two identical particles exists to the extent $\nu ( P_{\Psi(1, 2)} ; \epsilon)$ then the jth particle exists to the extent $\nu ( \rho_{0}(j) ; \epsilon )
= {1\over 2} \nu(P_{\psi_{+}(j)} ;  \epsilon)  + {1\over 2} \nu(P_{\psi_{-}(j)} ;  \epsilon)$.
Therefore the $ith$ coordinate qrumber value of the  jth particle's position is ${1\over 2}x^{i}_Q(\nu(P_{\psi_{+}(j)} ;  \epsilon)) + {1\over 2}x^{i}_Q(\nu(P_{\psi_{-}(j)} ;  \epsilon)) $.

In the EPR - type experiment, we used $\psi_{+} = u_L$ and $ \psi_{-} = v_R$ which are assumed to be non-overlapping narrow Gaussians, $u_L$ moving with speed $v$ to the left along the axis $\vec a$, $v_R$ moving with speed $v$ to the right along $\vec a$. Then the qrumber value of $\vec a \cdot \vec x(j)$ for the jth particle is approximately zero.

On the other hand when we measure $\vec a \cdot \vec x(j)$ for the jth particle in the vicinity of either of the states $P_{\psi_{+}(j)}$ or $P_{\psi_{-}(j)}$ we will get a value, which leads us to deduce that the $jth$ particle has a $\vec a \cdot \vec x(j)$ qrumber value to both the extents $\nu(P_{\psi_{+}(j)} ;  \epsilon)$ and  $\nu(P_{\psi_{-}(j)} ;  \epsilon)$.  It follows from the orthogonality of the wave functions and $\epsilon < 1$ that the extents are disjoint. 

The calculation for the two particles trajectories is based on the assumption that the particles are moving freely. Since we have assumed that for each particle the single particle wave functions are non-overlapping it is easy to show that the terms
$ Tr P_{\Psi(1,2)}({\hat P_{i}(1)\over m_1} \otimes \hat X_{j}(2) + \hat X_{i}(1) \otimes {\hat P_{j}(2)\over m_2})$, $ Tr P_{\Psi(1,2)} ( \hat X_{i}(1) \otimes \hat X_{j}(2))$ and $Tr P_{\Psi(1,2)} (\hat P_{i}(1)\otimes \hat P_{j}(2))$  vanish for all $i, j$ except for the component $(a,a)$. For instance, $Tr P_{\Psi(1,2)} (\hat X_{a}(1)\otimes \hat X_{a}(2)) = -(v.t)^2 $, where $X_{a} = \vec a \cdot \vec X/(\vec a \cdot \vec a)$. These results extend to the open set 
$W = \nu ( P_{\Psi(1, 2)} ; \epsilon)$ by continuity.

The extension of these results to an EPR-Bohm-Bell  experiment with two identical spin one-half massive particles is straightforward: the qrumber trajectories obtained for spinless particles in the EPR-experiment are split by the interaction between the spin and the magnetic field. The pair of qrumber trajectories become four qrumber trajectories.


\section{Acknowledgments}
Parts of this work has benefitted from comments and criticisms of members of the
Centre for Time at Sydney University, where JVC is an Honorary Associate. 
T.D. is a Postdoctoral Fellow of the Fonds voor
Wetenschappelijke Onderzoek, Vlaanderen and also thanks supports of
Inter-University Attraction Pole Program of the Belgian government
under grant V-18, International Solvay Institutes for Physics and Chemistry, the Concerted Research Action Photonics in
Computing and the research council (OZR) of the VUB. The work was realised during a 6 months visit of T.D. at Macquarie's university supported by a F.W.O. mobility grant.


%
%

%


\end{document}